\documentclass[aps,twocolumn]{revtex4-1}
\usepackage{graphicx}

\begin{document}

\title{Three-dimensional Brownian diffusion of rod-like macromolecules in the presence of randomly distributed
 spherical obstacles: Molecular dynamics simulation}

\author{Farzaneh Sakha}

\affiliation{Institute for Advanced Studies in Basic Sciences,
Department of Physics, Zanjan 45137-66731, Iran}

\author{Hossein Fazli}

\email{fazli@iasbs.ac.ir} \affiliation{Institute for Advanced
Studies in Basic Sciences, Department of Physics, Zanjan
45137-66731, Iran}
\affiliation{Institute for Advanced Studies in
Basic Sciences, Department of Biological Sciences, Zanjan
45137-66731, Iran}

\date{\today}

\begin{abstract}
Brownian diffusion of rod-like polymers in the presence of
randomly distributed spherical obstacles is studied using
molecular dynamics (MD) simulations. It is observed that
dependence of the reduced diffusion coefficient of these
macromolecules on the available volume fraction can be described
reasonably by a power law function. Despite the case of
obstructed diffusion of flexible polymers in which reduced
diffusion coefficient has a weak dependence on the polymer
length, this dependence is noticeably strong in the case of
rod-like polymers. Diffusion of these macromolecules in the
presence of obstacles is observed that is anomalous at short time
scales and normal at long times. Duration time of the anomalous
diffusion regime is found that increases very rapidly with
increasing both the polymer length and the obstructed volume
fraction. Dynamics of diffusion of these polymers is observed
that crosses over from Rouse to reptation type with increasing the
density of obstacles.

\end{abstract}

\maketitle

\section{Introduction}

The phenomenon of Brownian diffusion of particles in the presence
of fixed obstacles -- obstructed diffusion -- is very ubiquitous
in biology, chemistry, and physics. This phenomenon, as a
combination of several concepts such as obstructed random walk,
hydrodynamic interactions and topological hindrances, introduces
a challenging theoretical problem. Two simple models of this
problem namely three dimensional random walk in a regular lattice
with a fraction of its sites obstructed and diffusion of a
pointlike tracer in an ordered medium of spherical obstacles have
been studied theoretically and using lattice Monte Carlo method,
respectively \cite{Mackie,Mercier}. Also, existence of anomalous
diffusion regime in Brownian diffusion of particles due to the
presence of fixed and mobile obstacles has been reported
\cite{Saxton1,Saxton2,Wedemeier}.

In the context of obstructed diffusion, the other subject is the
diffusion of solute molecules in hydrogels for which there are
numerous suggestions (analytical, simulation, and empirical) for
diffusion coefficient as a function of obstacles density. In a
review of studies of this phenomenon it has been shown that
diffusion of solute molecules in hydrogels composed of rigid
(flexible) polymers can be described reasonably by obstruction
(hydrodynamic) models \cite{Amsden1}. Also, the effect of the
structure of obstacles medium on diffusion of fluid molecules has
been studied recently \cite{Sung}.

The other issue in this category is the Brownian diffusion of
polymeric macromolecules in a medium of obstacles which has
different branches depending on the polymer chains properties and
the strength of induced confinement. In diffusion of polymer
chains in a random media, such as exclusion chromatography, the
polymers enter in cavities of typical size comparable to or
larger than their size. In transition between these cavities the
chains have to overcome entropic barriers which are set up due to
reduction of their possible configurations
\cite{Muthukumar3,Muthukumar1,Muthukumar2,Yamakov2}. In a
semidilute solution of polymer chains or when a polymer chain
diffuses in a dense medium of fixed obstacles, typical size of
cavities is smaller than the polymer size and both topological
hindrances and entropic effects are of important role
\cite{Yamakov1,Avramova,Chang}. Ratio of the strength of these two
effects depends on the flexibility of the chains. The extreme
case of a stiff polymer chain is a rod-like polymer for which
internal entropy is negligible and topological interactions are
dominant in its diffusion in a crowded environment.

Despite pure translational Brownian motion of small isotropic
particles, Brownian motion of anisotropic macromolecules such as
rod-like polymers contains both translational and rotational
parts. Coupling of translational and rotational motions of such
macromolecules makes understanding and visualization of their
Brownian motion noticeably difficult \cite{Han}. Anisotropy
affects Brownian motion of a macromolecule more dramatically in
the presence of other macromolecules or fixed obstacles. In this
case, both hydrodynamic friction coefficient and confinement
effects due to the presence of other macromolecules or obstacles
are anisotropic. A rod-like macromolecule experiences the least
(the most) collisions with other macromolecules or obstacles when
it moves along (perpendicular to) its axis.

Numerous interesting phenomena are known to arise from special
shape of rod-like macromolecules. Nematic ordering in dense
solution of these macromolecules due to only excluded volume
interaction \cite{Onsager,Cinacchi}, a variety of their
orientational orderings resulted from the competition of entropic
effects with long- and short-range interactions
\cite{Fazli,Sarah}, and their rich dynamic behavior in crowded
environments \cite{Kob2,Tao,Hofling,Baskaran} are examples of
these phenomena.

In this paper we study Brownian diffusion of a spherical particle
and rod-like macromolecules in the presence of randomly
distributed spherical obstacles using molecular dynamics
simulations. We find that dependence of the reduced diffusion
coefficient of the spherical particle and rod-like macromolecules
on the available volume fraction can be described by a simple
power law function. In the case of rod-like macromolecules, the
exponent of mentioned power law dependence is an increasing and
saturating function of the macromolecule length. Despite the case
of flexible polymers in which reduced diffusion coefficient has a
weak dependence on the length of polymer \cite{Avramova},
noticeably strong dependence is observed here which originates
from stiffness of rod-like macromolecules. In obstructed diffusion
of these polymers, an anomalous diffusion regime is observed which
its duration time is a rapidly increasing function of both the
polymer length and the obstacles density. With increasing the
density of obstacles, long time diffusion of these rod-like
polymers crosses over from Rouse to reptation dynamics.

The rest of the paper is organized as follows. The model and the
simulation method are described in Sec. \ref{Model}. The results
are presented in Sec. \ref{Results}. Conclusions and a short
discussion are presented in Sec. \ref{Conclusion}.

\section{The model and the simulation method}\label{Model}

In our simulations which are performed with the MD simulation
package ESPResSo \cite{ESPResSo}, a polymer is modeled as a
bead-spring chain of length $N$ ($N$ spherical monomers of
diameter $\sigma$). Successive monomers of the polymer chain are
bonded to each other by a FENE (finite extensible nonlinear
elastic) potential \cite{Grest},
\begin{equation}
u_{bond}(r)=\left\lbrace
     \begin{array}{l l}
       - \frac{1}{2}k_{bond}R_{0}^{2}\ln(1-(\frac{r}{R_{0}})^{2})& \text{if $r<R_{0}$},\\
       \infty & \text{if $r \geq R_{0}$},
  \end{array}
\right. \label{FENE}
\end{equation}
with bond strength $k_{bond}=30\varepsilon/\sigma^{2}$ and maximum
bond length $R_{0}= 1.5\sigma$. Bending elasticity of the chain is
modeled by a bond angle potential,
\begin{equation}
u_{bend}(r)=k_{bend}(1-\cos\theta) \label{angle},
\end{equation}
in which $\theta$ is the angle between two successive bond vectors
and $k_{bend}$ is the bending energy of the chains. The value of
the chain persistence length relative to its contour length,
$\frac{l_p}{L_c}$, which is a measure of the flexibility of the
chain depends on the value of $k_{bend}$ as
$l_p=\frac{k_{bend}}{k_BT}\sigma$. To model a rod-like polymer we
consider $l_p\simeq 100 L_c$ in our simulations. We also model
the spherical obstacles by fixed spheres of diameter $\sigma$
which are randomly distributed inside the simulation box with
their overlapping permitted. Excluded volume interaction between
monomers and the obstacles is modeled by a shifted Lennard-Jones
potential,
\begin{equation}
 u_{LJ}(r) = \left\lbrace
  \begin{array}{l l}
    4\varepsilon\
    \{(\frac{\sigma}{r})^{12}-(\frac{\sigma}{r})^{6}+\frac{1}{4}\} & \text{if $r<r_{c}$},\\
    0 & \text{if $r \geq r_{c}$},
  \end{array}
\right. \label{slj}
\end{equation}
in which $\epsilon$ and $\sigma$ are the usual Lennard-Jones
parameters and the cutoff radius is $r_{c}=2^{1/6} \sigma$.
Average distance between neighboring monomers which results from
combination of FENE and Lennard-Jones interactions is $b\simeq
0.97 \sigma$. Simulation box is cubic with periodic boundary
conditions and the temperature is kept fixed at
$k_{B}T=1.0\varepsilon$ using a Langevin thermostat. 
Simulation box length is $L=20 \sigma$ and $L=25 \sigma$ in
simulations of polymers with $N\leq 17$ and the polymer with $N=21$, respectively.
MD time step in our simulations is $\tau=0.01\tau_0$ in which
$\tau_0=\sqrt{\frac{m\sigma^2}{\varepsilon}}$ is the MD time
scale and $m$ is the mass of the monomers. Dynamics of the
monomers is assumed to be overdamped by using the friction
coefficient $\Gamma=\frac{\epsilon \tau_0}{\sigma^2}$ for each
monomer in the Langevin thermostat. In fact, we
model the solvent implicitly by the stochastic and dissipation terms in Langevin
equation instead of adding solvent molecules explicitly to the system.
Accordingly, hydrodynamic interactions are not considered here and 
dynamics of polymer chains in the absence of obstacles is of 
Rouse type. In each simulation we first
randomly distribute the obstacles inside the simulation box in a
way that their overlapping is permitted. Then we put the rod-like
polymer in a random position with a random orientation inside the
box and start the simulation. After equilibration of the system
which is needed because of used potentials and fixed temperature,
we probe center of mass position, orientation, and velocity of
the polymer for a long time ($N_t$ time steps) to calculate
correlation functions. From recorded position of a diffusing
particle (position of the center of mass in the case of rod-like
polymers), mean square displacement (MSD) at time $t=n_t\tau$ is
calculated as
\begin{equation}
MSD=\frac{1}{N_t-n_t}\sum_{n_s=0}^{N_t-n_t}|\vec{r}((n_s+n_t)\tau)-\vec{r}(n_s\tau)|^2,
 \label{MSD}
\end{equation}
in which $\vec{r}(t)$ is the position vector of the particle at
time $t$. Diffusion coefficient for each realization of obstacles
is obtained from time dependence of MSD. For fixed values of the
polymer length and obstacles density we repeat our simulations
with 10 different realizations of obstacles to average quantities
and obtain error bars.

\section{Results}\label{Results}
\begin{figure}[t]
\includegraphics[width=.9 \columnwidth]{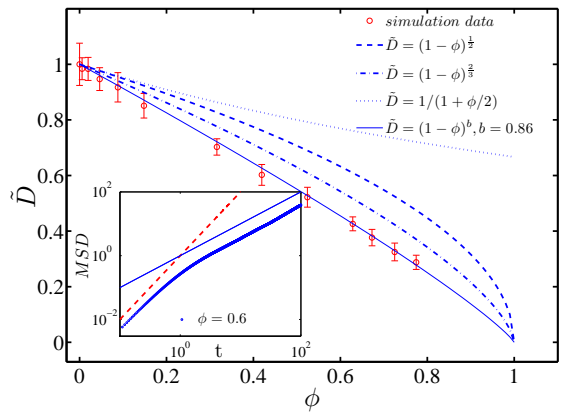}
\caption{(Color online) Reduced diffusion coefficient of a
spherical particle in the presence of randomly distributed
spherical obstacles as a function of obstructed volume fraction.
Some suggested functions for dependence of diffusion coefficient
on obstructed volume fraction are also plotted. Simulation data
fits well with a simple power law function of available volume
fraction with exponent $b=0.86$. Inset: A sample simulation data
for particle MSD versus time in logarithmic scales. Ballistic and
diffusion regimes can be seen (slopes of dashed and solid lines
are 2.0 and 1.0, respectively).} \label{fig1}
\end{figure}

To study obstructed diffusion of a particle or a macromolecule,
we first obtain its diffusion coefficient in the absence of
obstacles ($D_0$) at given temperature, $T$, and friction
coefficient, $\Gamma$, which is in consistence with Einstein
relation, $D_0=\frac{k_BT}{\Gamma}$. Then we calculate diffusion
coefficient, $D$, in the presence of fixed spherical obstacles of
given density. Obstructed volume fraction, $\phi$, is calculated
using ESPResSo package (a mesh based cluster algorithm
\cite{ESPResSo}). This calculation is checked and justified with
examples for which obstructed volume fraction can be calculated
exactly. Reduced diffusion coefficient for each value of $\phi$
is defined as $\tilde{D}=\frac{D}{D_0}$.
\begin{figure}[t]
\includegraphics[width=1. \columnwidth]{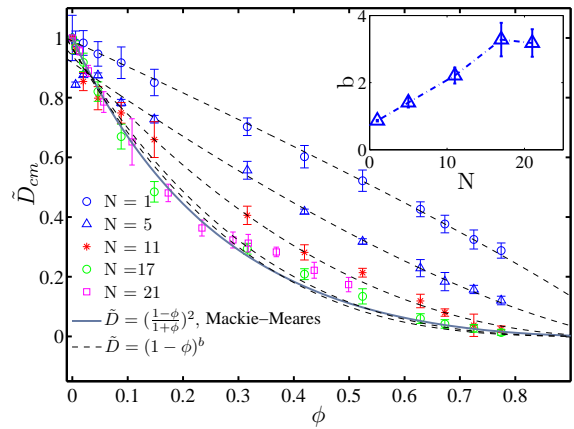}
\caption{(Color online) Reduced diffusion coefficient of the
center of mass of rod-like polymers of different length versus
obstructed volume fraction, $\phi$. Data for spherical particle,
$N=1$, is shown again for comparison. Power low function
$\tilde{D}_{cm}=(1-\phi)^b$ can be fitted to simulation data for
all values of $N$ reasonably (dashed lines). Mackie-Meares
function is also plotted by solid line (see Sec. \ref{rods} in the
text). Inset: The exponent $b$ versus polymer length, $N$, that is
an increasing semi-linear function showing saturation for $N\geq
17$.} \label{fig2}
\end{figure}

\subsection{Obstructed diffusion of a spherical
particle}\label{spheres}
 From MD simulations of the Brownian
diffusion of a spherical particle in the presence of fixed
spherical obstacles, we calculate diffusion coefficient at
different values of obstacles density. Diffusion coefficient is
calculated from the plot of the particle MSD versus time when its
dynamics becomes of diffusion type at long enough times (see the
inset of Fig. \ref{fig1}). Reduced diffusion coefficient of the
spherical particle versus obstructed volume fraction, averaged
over 10 realizations of the system for each value of $\phi$ is
shown in Fig. \ref{fig1}. Examples of functions suggested for
$\phi$ dependence of $\tilde{D}$ such as
$\tilde{D}=(1-\phi)^{1/2}$ form Ref. \cite{Ilgenfritz},
$\tilde{D}=1/(1+\phi/2)$ from Ref. \cite{Mercier} and
$\tilde{D}=(1-\phi)^{2/3}$ form Ref. \cite{Mittal} are also shown
in this figure. As it can be seen, the best fitted function is a
power law function of available volume, $1-\phi$, with an
exponent $b=0.86$. The next closer function is the prediction of
available volume theory \cite{Mittal}. Our data points are
considerably far from that of Ref. \cite{Mercier} at high values
of $\phi$. As we checked by running the simulation with the same
configuration of the obstacles as of Ref. \cite{Mercier}, this
difference doesn't originate from obstacles configuration. This
difference may be because of the fact that studying the diffusion
phenomena in the presence of obstacles by lattice and
continuous-space methods gives different results \cite{Sung}.
Note that the difference between our results and those of Ref.
\cite{Mittal} originates from the fact that we have not used
explicit fluid particles in our simulations. $D_0$ in
our work is constant and does not depend on fluid density or free
volume fraction. In the case of Ref. \cite{Mittal} however, $D_0$
depends on the fluid density (Eq. 6 of Ref. \cite{Mittal}).

\subsection{Brownian diffusion of rod-like macromolecule in the
presence of spherical obstacles}\label{rods}

\begin{figure}[t]
\includegraphics[width=1. \columnwidth]{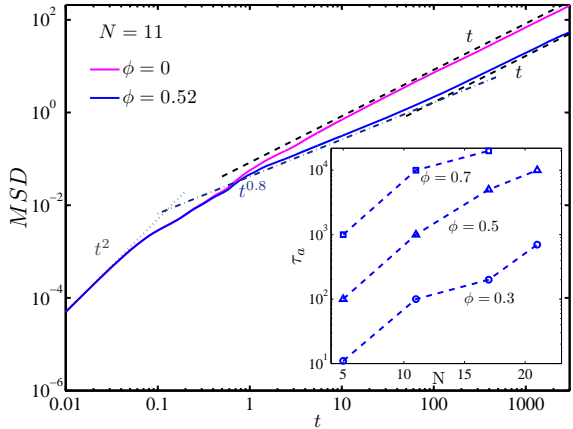}
\caption{(Color online) Log-log plot of time-dependent center of
mass MSD for a polymer of length $N=11$ at $\phi=0$ and
$\phi=0.52$. As it can be seen, despite the $\phi=0$ case an
anomalous diffusion regime at intermediate times is obviously
seen in the case of $\phi=0.52$. In time interval between
ballistic and diffusion regimes (approximately from $t=0.1$ to
$t=1$) velocity autocorrelation function is not yet vanished.
Duration time of anomalous diffusion regime is observed that
increases with increasing the polymer length and the value of
$\phi$. Inset: Duration time of anomalous diffusion regime versus
the polymer length at three different obstructed volume
fractions. Simulation of the polymer with $N=21$ monomers at
$\phi=0.7$ takes very long time and we had not a reasonable data
for these parameters. } \label{fig3}
\end{figure}

In simulations of the Brownian motion of rod-like polymers in the
presence of spherical obstacles we use Rouse dynamics.
Accordingly, hydrodynamic interactions are not considered and in
the absence of obstacles friction coefficient is not direction
dependent. Anisotropy in dynamics of these macromolecules
originate solely from obstruction effects of the obstacles. After
equilibration of the system for each value of obstructed volume
fraction, $\phi$, we record the position of the polymer center of
mass and its orientation (a unit vector parallel to the polymer
axis) at all simulation time steps. From trajectory of the center
of mass and time dependence of polymer orientation we calculate
diffusion coefficient and probe orientational relaxation of the
polymer. Reduced diffusion coefficient of the center of mass
versus $\phi$ for four polymers of different length are shown in
Fig. \ref{fig2}. These diffusion coefficients are obtained from
time dependent MSD of the polymers center of mass, two samples of
which are shown in Fig. \ref{fig3}. Considering that
functionality of dependence of the diffusion coefficient on the
obstructed volume fraction is an important question in the
subject of obstructed diffusion, we search this functionality
here. Our results show that this dependence in the case of
rod-like polymers as well as spherical particle can be reasonably
described by a power law function of available volume fraction,
$(1-\phi)^b$, in which the value of the exponent $b$ depends on
the length of the polymer. It has already been suggested that
obstructed diffusion coefficient can be written as a
function of a single parameter, namely the available volume
fraction \cite{Mittal}. As it is shown in the inset of Fig.
\ref{fig2}, $b$ grows linearly with increasing the polymer
length, $N$, and becomes saturated. The function for $\phi$
dependence of $\tilde{D}$ suggested by Mackie and Meares
\cite{Mackie},
\begin{equation}
\tilde{D}=(\frac{1-\phi}{1+\phi})^2,
 \label{Mackie}
\end{equation}
is also plotted as solid line in Fig. \ref{fig2}. This function is
quite far from spherical particle and short polymers data. It is
close to the data of long polymers of lengths $N=17$ and $N=21$.
Noticeable dependence of $\tilde{D}_{cm}$ on polymer length which
can be seen in this figure is quite different from the case of
flexible polymers which is reported in ref. \cite{Avramova}. In a
crowded environment that diffusion dynamics of individual
monomers is slow relative to the case with no obstacles,
diffusion coefficient of a flexible polymer is also diminished by
the same factor as monomers diffusion coefficient. In this case,
the polymer reduced diffusion coefficient becomes independent of
its length. In the case of rod-like polymers however, the
monomers have to obey whole polymer dynamics because of its
rigidity. In this case longer polymers experience stronger
obstruction because of stronger spatial hindrance and the polymer
diffusion coefficient considerably depends on the polymer length.
Dependence of flexible polymers diffusion coefficient in
disordered porous material on matrix volume fraction when
numerous polymers diffuse simultaneously has been already
reported \cite{Chang}.

In the plot of center of mass MSD versus time in logarithmic
scales, Fig. \ref{fig3}, at $\phi=0$ in addition to ballistic and
diffusion regimes which correspond to linear parts of slopes 2
and 1, respectively, a third semi linear region can be seen. In
this region velocity autocorrelation function is not yet vanished
and the diffusion regime is not yet started. As it is shown in
Fig. \ref{fig3}, another linear regime in addition to ballistic
and diffusion regimes can be seen in which velocity
autocorrelation function is vanished but the slope of fitted line
differs from unity. This region corresponds to anomalous diffusion
of the polymer over a time interval during its dynamics. Our
results show that duration of anomalous diffusion of the rod-like
polymer increases rapidly with increasing its length and the
value of the obstructed volume fraction, $\phi$. In the inset of
Fig. \ref{fig3} time length of anomalous diffusion regim of the polymer,
$\tau_a$, versus its length at three different values of $\phi$
are shown. These times are roughly obtained from our MSD data as
the time interval in which data points are clearly out of two
lines of slopes 2 and 1 and the semilinear part in which velocity
autocorrelation function is not yet vanished (see main part of
Fig. \ref{fig3}). As it can be seen, the value of $\tau_a$
changes its order of magnitude with increasing the strength of
obstruction, namely the value of $\phi$ and the length of the
polymer, $N$. Increasing of the duration of anomalous diffusion regime with
increasing the obstructed volume fraction has also been reported
already in the context of particle diffusion in quenched media
(see for example refs. \cite{Saxton1,Saxton2}).
\begin{figure}[t]
\includegraphics[width=1. \columnwidth]{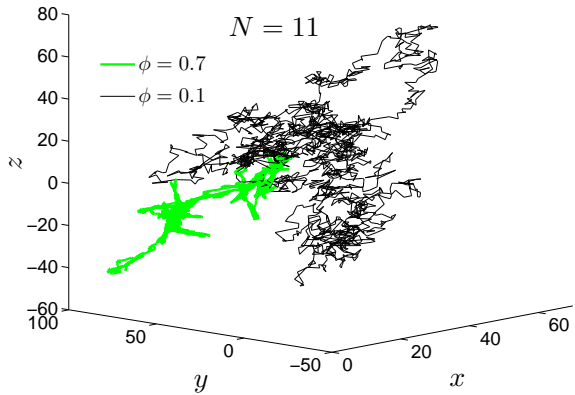}
\caption{(Color online) Sample trajectory of the center of mass
of a rod-like polymer with $N=11$ monomers at two obstructed
volume fractions $\phi=0.1$ and $\phi=0.7$ during $\Delta
t=5\times 10^6$ MD time steps. Trajectory in the case $\phi=0.7$
is more anisotropic and contains linear parts showing that the
polymer dynamics is of reptation type. } \label{fig4}
\end{figure}

\begin{figure}[t]
\includegraphics[width=1. \columnwidth]{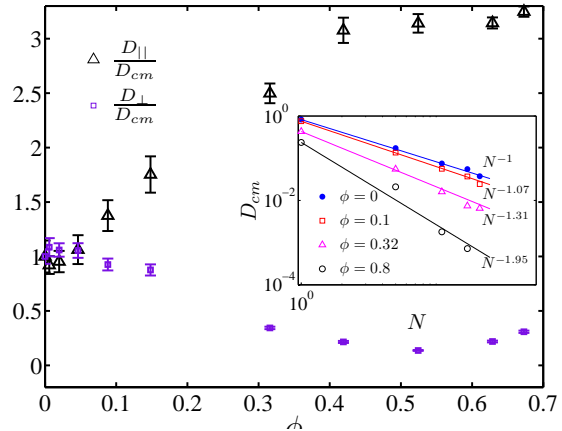}
\caption{(Color online) Diffusion coefficient of a rod-like
polymer of length $N=17$ along $(D_{||})$ and perpendicular to
$(D_{\perp})$ its axis versus $\phi$. As it can be seen, $D_{||}$
and $D_{\perp}$ are respectively increasing and decreasing
functions of $\phi$ showing that the polymer dynamics crosses
over to the reptation type. Inset: Log-log plot of $D_{cm}$
versus polymer length, $N$, which shows that with increasing the
fraction of obstructed volume, the exponent $\alpha$ in
$D_{cm}\sim N^{-\alpha}$ changes from 1 (Rouse dynamics) to
$\simeq 2$ (reptation dynamics). } \label{fig5}
\end{figure}

To investigate anisotropy in diffusion dynamics of rod-like
polymers in strongly obstructed conditions, a sample of which is
shown in Fig. \ref{fig4}, we calculate diffusion coefficients
$D_{||}$ and $D_{\perp}$ in directions parallel and perpendicular
to the polymer axis, respectively. It is done by decomposing
displacement of the polymer center of mass in each time step into
two parts, parallel and perpendicular to its axis. For a rod-like
diffusing object it is known that
\begin{equation}
D_{cm}=\frac{D_{||}+2D_{\perp}}{3}
 \label{Mackie}
\end{equation}
and in the reptation regime where $D_{\perp}\simeq 0$, this
relation becomes $D_{cm}\simeq\frac{1}{3}D_{||}$. For a rod-like
polymer consisting of $N=17$ monomers, $D_{||}$ and $D_{\perp}$
are shown as functions of $\phi$ in Fig. \ref{fig5}. With
increasing the value of $\phi$, it can be seen that $D_{\perp}$
vanishes meaning that the polymer dynamics becomes of reptation
type. It is known in polymer physics \cite{Rubinstein} that
despite the Rouse dynamics in which $D_{cm}\sim N^{-1}$, in the
reptation regime, $D_{cm}\sim N^{-2}$. In the inset of Fig.
\ref{fig5}, log-log plot of $D_{cm}$ versus polymer length, $N$,
is shown. Crossing over from Rouse to reptation dynamics with
increasing $\phi$ is obviously seen in this figure.

\begin{figure}[t]
\includegraphics[width=1. \columnwidth]{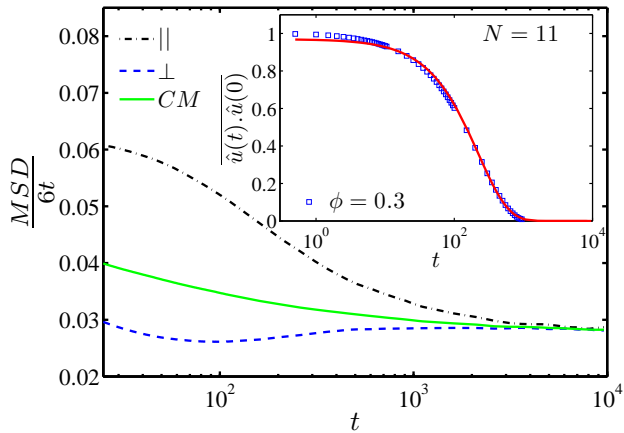}
\caption{(Color online) Relaxation of the diffusion coefficients
along and perpendicular to initial direction of the polymer to
the value of its center of mass diffusion coefficient as time
elapses. Inset: Autocorrelation function of the unit vector along
the polymer axis (defined in the text) versus the time for the
same polymer and fitted function, $A\exp(-t/\tau_r)$, with
$\tau_r\simeq 225$. } \label{fig6}
\end{figure}

To investigate orientational relaxation of a rod-like polymer
which is related to duration time of their anomalous diffusion
regime, we calculate autocorrelation function of the unit vector
corresponding to its direction, $\hat{u}(t)$, defined as
\begin{equation}
\overline{\hat{u}(t).\hat{u}(0)}=\frac{1}{N_t-n_t}\sum_{n_s=0}^{N_t-n_t}\hat{u}((n_s+n_t)\tau).\hat{u}(n_s\tau).
\label{Auto}
\end{equation}
Relaxation time, $\tau_r$, is calculated by fitting the
exponential function, $A\exp(-t/\tau_r)$, to the simulation data
as is shown in the inset of Fig. \ref{fig6} for a polymer of
length $N=11$ at $\phi=0.3$. Also, following Ref. \cite{Han}, we
probe relaxation of diffusion coefficients along and
perpendicular to the polymer initial direction from their initial
values to the value of $D_{cm}$ over the time. At small times,
diffusion coefficient along the polymer is larger than that in
perpendicular direction. As time goes on and the polymer forgets
its initial direction, diffusion coefficients parallel and
perpendicular to the initial direction become of the same value
(see Fig. \ref{fig6}). As the figure shows, relaxation time
obtained from both methods are approximately the same. According
to our results, for a polymer of given length, the value of the
orientational relaxation time is comparable to the time length of
its anomalous diffusion regime. It shows that the existence of
anomalous diffusion regime comes from anisotropy in
the shape of the polymer. A rod-like polymer experiences ordinary
diffusion regime only at time scales larger than its
orientational relaxation time. For a long polymer at high values
of obstructed volume fraction this relaxation time could be very
long and may be longer than simulation or experiment time.

\section{Conclusions and discussion}\label{Conclusion}

In conclusion, MD simulation study of the Brownian diffusion of a
particle and rod-like polymers in the presence of randomly
distributed spherical obstacles showed that: 1- Obstructed volume
fraction dependence of the reduced diffusion coefficient of a
spherical particle, obtained from our off lattice MD simulations
differs from that of lattice based study of obstructed random
walk. 2- Reduced diffusion coefficient of both rod-like polymers
and spherical particle have a power law dependence on the
available volume fraction. The exponent of the power law function
in the case of rod-like polymers is an increasing and saturating function of the
polymer length. 3- Despite the case of flexible polymers for
which a weak dependence of the reduced diffusion coefficient on
the obstructed volume fraction has been reported, this dependence
is noticeably strong in the case of rod-like polymers. 4-
Obstructed diffusion of rod-like polymers is observed that is
anomalous at short time scales. Time length of the anomalous
diffusion regime is of the same order of the polymer
orientational relaxation time and is a rapidly increasing
function of both the polymer length and the obstructed volume
fraction. 5- With increasing the density of obstacles, diffusion
dynamics of rod-like polymers is observed that crosses over from
Rouse to reptation type.

Although MD simulation of rod-like polymers as presented here is
too time consuming because of using numerous potentials, such
simulations doesn't suffer from dependence of results on the
selected lattice and can give better results. Specially, with
overdamped dynamics of diffusing particle and working with
constant temperature, the obstruction effect of the obstacles is
considered very well.

The main difference between dynamics of flexible polymers and
that of semiflexible or rod-like polymers comes from anisotropy
in the shape of stiff polymers. Clearly the effects of
anisotropic shape of a macromolecule on its dynamics becomes more
intense in the presence of obstacles. Dynamics of a flexible
polymer in the presence of obstacles is dominantly affected by
obstruction of monomers dynamics alone. However, in the case of
stiff polymers in addition to obstruction effect of the obstacles
on the monomers, topological hindrance originated from polymer
stiffness also plays an important role. Strong dependence of
diffusion coefficient of rod-like polymers on the obstructed
volume fraction relative to flexible polymers and existence of
anomalous diffusion regime in dynamics of these polymers are
consequences of above mentioned difference.

Analytical approach to obstructed diffusion of rod-like polymers
seems a challenging problem. Taking into account the size of the
obstacles and non zero diameter of the polymer in obtaining
dependence of diffusion coefficient on obstructed volume fraction
is not straight forward and an easy task.

\acknowledgements The authors gratefully acknowledge support by
the Institute for Advanced Studies in Basic Sciences (IASBS)
Research Council under grant No. G2009IASBS134. We also would like
to acknowledge Sharareh Tavaddod for useful discussions.

\end{document}